\documentclass{article} 
  
\usepackage{amssymb,eufrak,epsfig,latexsym} 
\input tcilatex 
\newenvironment{beweis}{{\bfseries\upshape\selectfont Proof:}} 
 {\ifmmode\rlap{\hskip\displaywidth\rule{2mm}{2mm}}\else\hfill $\rule{2mm}{2mm}$\fi} 
\def\ABB#1#2#3{\vskip6mm\def\Test{#2}\ifx\Test\empty\centerline{%
 \epsfig{figure=#1}}\else\centerline{\epsfig{figure=#1, width=#2}}\fi%
 \def\Test{#3}\ifx\Test\empty\else\vskip0mm\global\advance\c@Abbildung %
 by 1\def\@currentlabel{(\thechapter-\theAbbildung)}%
 \centerline{\mbox{\rmfamily\mdseries\upshape\fontsize{9pt}{13pt}%
 \selectfont \begin{tabular}{rl} Abbildung (\thechapter-\theAbbildung):%
 \kern -3mm & #3\end{tabular}}}\fi\vskip5mm}

\begin{document} 

\title{On future asymptotics of polarized Gowdy  
$\mathbb T^3$-models} 
\author{Thomas Jurke\\
{\em Max-Planck-Institut f\"ur Gravitationsphysik}\\ 
{\em Am M\"uhlenberg 1}\\ 
{\em 14476 Golm}\footnote{current address: 
Weierstra{\ss}-Institut 
f\"ur Angewandte Analysis und 
Stochastik, Mohrenstra{\ss}e 39, 10117 Berlin}}
\maketitle 

\begin{abstract} 
Gowdy's model of cosmological spacetimes is a much investigated 
subject in classical and quantum gravity. Depending on spatial 
topology re\-collapsing as well as expanding models are known. 
Several analytic tools were used in order to clarify singular 
behaviour in this class of spacetimes. Here we investigate the 
structure of a certain subclass, the polarized Gowdy models with 
spatial $\mathbb T^3$-topology, in the large. 
The a\-sym\-pto\-tics for general solutions of the dynamical equation for one 
of the gravitational degrees of freedom plays a key role 
while the asymptotic behaviour of the remaining metric function 
is a result of solving the Hamiltonian constraint equation. 
Using both we are able to prove (future) geodesic 
completeness in all spacetimes of this type. 
\end{abstract}

\section{Introduction} 
%\begin{eqnarray} 
%\frac{\partial^2}{\partial t^2}W\kern1mm+\kern1mm
%\frac1t\frac{\partial}{\partial t}W\kern1mm-\kern1mm
%\frac{\partial^2}{\partial x^2}W&=&\sinh W\cosh W
%\left(\left(\frac{\partial Q}{\partial t}\right)^2-
%\left(\frac{\partial Q}{\partial x}\right)^2\right)
%	\label{GowdyW}\\ 
%\frac{\partial^2}{\partial t^2}Q\kern1mm+\kern1mm
%\frac1t\frac{\partial}{\partial t}Q\kern1mm-\kern1mm
%\frac{\partial^2}{\partial x^2}Q&=&-2\coth W
%\left(\frac{\partial Q}{\partial t}\frac{\partial W}{\partial t}-
%\frac{\partial Q}{\partial x}\frac{\partial W}{\partial x}\right)
%        \label{GowdyQ}\\ 
%\frac{\partial^2}{\partial t^2}a\kern1mm+\kern1mm
%\frac1t\frac{\partial}{\partial t}a\kern1mm-\kern1mm
%\frac{\partial^2}{\partial x^2}a&=&\frac12
%\left(\left(\frac{\partial W}{\partial x}\right)^2+\sinh^2 W
%\left(\frac{\partial Q}{\partial x}\right)^2\right)
%        \label{Gowdya}
%\end{eqnarray}
%\begin{eqnarray}
%\frac{\partial}{\partial t}a&=&
%-\frac1{4t}+\frac{t}4\left[
%\left(\frac{\partial W}{\partial t}\right)^2+
%\left(\frac{\partial W}{\partial x}\right)^2+
%\right.\nonumber\\
%&&\left.
%\sinh^2 W\left(
%\left(\frac{\partial Q}{\partial t}\right)^2+
%\left(\frac{\partial Q}{\partial x}\right)^2
%\right)
%\right]
%	\label{Gowdyat}\\ 
%\frac{\partial}{\partial x}a&=&\frac{t}2\left(
%\frac{\partial W}{\partial t}
%\frac{\partial W}{\partial x}+
%\sinh^2 W\frac{\partial Q}{\partial t}
%\frac{\partial Q}{\partial x}
%\right)		
%	\label{Gowdyax} 
%\end{eqnarray} 
The cosmological spacetime model by Gowdy (1974) has experienced a great
deal of attention over the last years. These models are especially 
interesting because they allow rigorous analytical investigations in
inhomogeneous spacetimes. Various spatial topologies are allowed but~--~as
is mostly done~--~we will concentrate on vacuum spacetimes with $\mathbb
R_+\times\mathbb T^3$-topology. A basic analytical tool for this class of
spacetimes was provided by Moncrief~(1981) who proved existence of global
smooth solutions for the corresponding Einstein equations. Investigations 
concerning the singular asymptotic behaviour were subjects of interest
almost exclusively. Now, we have a quite clear picture what happens
in the polarized as well as the full Gowdy $\mathbb T^3$-model (see
Isenberg, Moncrief~(1990), Rendall~(2000), Ringstr\"om~(2002) and references 
therein). On the other hand, due to Chru\'sciel, Isenberg and Moncrief~(1990)
there exist some statements about the asymptotic behaviour for large times in 
the case of polarized spacetimes. In this paper we prove such theorems, 
describing the long time asymptotics for general polarized Gowdy models
and using this results to prove future completeness of any causal geodesic.

\section{The polarized model}  
By definition a smooth, orientable spacetime is a cosmological model of Gowdy
type if it is a maximally extended, globally hyperbolic solution of the
vacuum Einstein equations having compact Cauchy surfaces and a 
pseudo-Riemannian metric, invariant under an effective $U(1)\times U(1)$ 
group action on these Cauchy surface. This action is generated by two
spacelike Killing vector fields which are assumed to have vanishing twist
constants. A Gowdy model is called polarized if the defining Killing vector
fields are everywhere mutually orthogonal. By virtue of this definition
we may describe a polarized Gowdy metric by the line element
\begin{eqnarray}
{\rm d}s^2&=&{\rm e}^{2a(t,x)}\left(-{\rm d}t^2+{\rm d}x^2\right)\kern1mm+
\kern1mmt\kern.5mm\left({\rm e}^{W(t,x)}{\rm d}y^2+{\rm e}^{-W(t,x)}
{\rm d}z^2\right).
\end{eqnarray}
The corresponding vacuum Einstein equations are
\begin{eqnarray} 
0&=&\frac{\partial^2}{\partial t^2}W\kern1mm+\kern1mm 
\frac1t\frac{\partial}{\partial t}W\kern1mm-\kern1mm 
\frac{\partial^2}{\partial x^2}W\label{Darboux}\\[2mm] 
0&=&\frac{\partial^2}{\partial t^2}a\kern1mm-\kern1mm 
\frac{\partial^2}{\partial x^2}a 
\kern1mm-\kern1mm\frac1{4t^2}\kern1mm+\kern1mm\frac14 
\left[\left(\frac\partial{\partial t}W\right)^2\kern1mm-\kern1mm 
\left(\frac\partial{\partial x}W\right)^2 
\right]\label{Wellena}\\[2mm] 
0&=& 
\frac2t\frac{\partial}{\partial x}a\kern1mm-\kern1mm 
\frac{\partial}{\partial t}W\kern1mm\frac{\partial}{\partial x}W 
\label{Impuls1}\\[2mm] 
0&=& 
\frac2t\frac{\partial}{\partial t}a\kern1mm+\kern1mm 
\frac1{2t^2}\kern1mm-\kern1mm\frac12\left[\left( 
\frac{\partial}{\partial t}W\right)^2\kern1mm+\kern1mm 
\left(\frac{\partial}{\partial x}W\right)^2\right]. 
\label{Hamilton1} 
\end{eqnarray} 
As in the general non-polarized case equation~(\ref{Wellena}) is always 
satisfied as 
a consequence of the others. Thus, we have essentially to deal with a single 
linear equation of second order while  
function~$a$ is the integral of the constraint  
equation~(\ref{Hamilton1}). 
  
\section{On the asymptotics for solutions of the  
Euler-Poisson-Darboux equation} 
 
Now, we start by investigating equation~(\ref{Darboux}). As in 
literature we will sometimes call this equation 
``Euler-Poisson-Darboux equation''.  
Following Moncrief~(1981) it will be sufficient to have only 
functions $W=W(t,x)$ of a certain regularity in mind
which are defined for all 
$(t,x)\in\mathbb R_+\times\mathbb T^1$. 
  
As a first result we get asymptotic homogenity for all 
solutions of~(\ref{Darboux}). Namely, we will show 
that for large~$t$ the mean value, 
\begin{eqnarray} 
\bar W(t) &\mathop{=}\limits_{\rm Df}& \frac1{2\pi} 
\intop\limits_0^{2\pi} W(t, x)\kern1mm {\rm d} x, 
\label{W_mittel} 
\end{eqnarray}  
will dominate while the $x$-dependent terms show a  
$t^{-\frac12}$ fall-off behaviour. 
 
\begin{theorem} 
Let $W\in\mathcal C^2(\mathbb R_+\times\mathbb T^1;\mathbb R)$  
be a solution of Euler-Poisson-Darboux equation~(\ref{Darboux}) 
and $\bar W$ as decleared in~(\ref{W_mittel}). Then $\bar W$ 
solves~(\ref{Darboux}), too. 
\end{theorem} 
 
\begin{beweis} 
For every $t\in\mathbb R_+$ function $W$ as well as its first and 
second derivative are continuous, hence 
$\mathbb T^1$-integrable and $\bar W$ is well-defined.  
Since~$W$ is sufficiently smooth $\bar W$ depends on its 
parameter~$t\in\mathbb R_+$ smoothly, hence
$\bar W\in\mathcal C^2(\mathbb R_+;\mathbb R)$ and 
\begin{eqnarray*}  
\frac{\rm d^2}{{\rm d}t^2}\bar W(t)&=&\frac1{2\pi}  
\frac{\rm d}{{\rm d}t}\int_0^{2\pi}\frac\partial{\partial t}  
W(t,x)\kern1mm{\rm d}x\kern2mm=\kern2mm  
\frac1{2\pi}\int_0^{2\pi}\frac{\partial^2}{\partial t^2}  
W(t,x)\kern1mm{\rm d}x.  
\end{eqnarray*} 
Now it is easy to integrate~(\ref{Darboux})  
\begin{eqnarray*} 
\frac1{2\pi} 
\intop\limits_0^{2\pi} 
\frac{\partial^2}{\partial t^2}W \kern1mm {\rm d} x 
\kern2mm+\kern2mm 
\frac1{2\pi t} 
\intop\limits_0^{2\pi} 
\frac{\partial }{\partial t}W\kern1mm {\rm d} x &=& 
\frac1{2\pi} 
\intop\limits_0^{2\pi} 
\frac{\partial ^2}{\partial x^2}W\kern1mm {\rm d} x\\[2mm] 
%\end{eqnarray*} 
%\begin{eqnarray*} 
\frac{\rm d^2}{{\rm d} t^2}\bar W \kern2mm+\kern2mm 
\frac1t\frac{\rm d }{{\rm d} t}\bar W&=& 
\frac1{2\pi}\left. 
\frac{\partial W}{\partial x}\right|^{2\pi}_0 
\end{eqnarray*} 
and since the derivative of smooth periodic functions have same 
periods we get 
\begin{eqnarray*} 
\frac{\partial^2}{\partial t^2}\bar W \kern2mm+\kern2mm 
\frac1t\frac{\partial }{\partial t}\bar W&=&0\kern2mm 
\mathop{=}\limits^! \kern2mm 
\frac{\partial^2}{\partial x^2}\bar W. 
\end{eqnarray*} 
\end{beweis} 
 
The general solution of~(\ref{Darboux}) which does not depend on $x$
is 
\begin{eqnarray} 
\bar W(t)&=&\gamma\kern1mm+\kern1mm 
\beta\cdot\ln t, \label{Whom} 
\end{eqnarray} 
where $\beta$ and $\gamma$ may be fixed by some initial values 
of~$\bar W$ and~${\bar W}_t$ at $t_0>0$. Due to the 
linearity of~(\ref{Darboux}) $\psi$ in
\begin{eqnarray*} 
t^{-\frac12}\psi(t,x)&\mathop{=} 
\limits_{\rm Df}& W(t,x)-\bar W(t) 
\end{eqnarray*} 
is unique and solves (\ref{Darboux}), too. Obviously this 
newly defined function~$\psi$,   
$\psi\in\mathcal C^2(\mathbb R_+\times\mathbb T^1;\mathbb R)$,  
has zero mean value for every~$t\in\mathbb R_+$  
\begin{eqnarray} 
\bar\psi(t)&\mathop{=}\limits_{\rm Df}& 
\frac1{2\pi}\intop\limits_0^{2\pi}\psi(t,x)\kern1mm{\rm d}x 
\kern2mm=\kern2mm0\label{mittellos} 
\end{eqnarray} 
and satisfies 
\begin{eqnarray} 
\frac{\partial^2}{\partial t^2}\psi \kern1mm-\kern1mm 
\frac{\partial ^2}{\partial x^2}\psi&=& 
-\frac1{4t^2}\psi. 
\label{Wellenpsi} 
\end{eqnarray} 
  
Now, in order to prove that any solution of this equation has 
to be bounded we need the following estimate as prerequisite.  
Here we have formulated the lemma in dependence of a 
parameters~$t\in\mathbb R_+$, since this is the form we will need. 
\begin{lemma}  
Let \label{Hilfssatz2}  
$f\colon{\mathbb R}_+\times {\mathbb T}^1\to {\mathbb R}$ be 
continuously differentiable and  
\begin{eqnarray*} 
\frac1{2\pi}\int_0^{2\pi} f(t,x)\kern1mm {\rm d} x&=&0 
\end{eqnarray*} 
for every $t$. Then we have for each $(t,x)\in\mathbb R_+ 
\times\mathbb T^1:$
\begin{eqnarray*}  
|f(t,x)|^2 &\le& 2\pi\intop\limits_0^{2\pi}\left|  
\frac\partial{\partial x} f(t,x)  
\right|^2\kern1mm {\rm d} x\kern2mm\equiv\kern2mm 2\pi\kern1mm 
\left|\left| f_x(t,.)\right|\right|_{L^2}^2  
\end{eqnarray*}  
\end{lemma}  
  
\begin{beweis}  
The mean value theorem of integral calculus for continuous 
functions gives us to any~$t$ some 
$x_0 \in {\mathbb T}^1$ with $f(t, x_0)=0$. Furthermore   
$|x-x_0|\le 2\pi$ is true for each $x\in\mathbb T^1$. Using 
Schwartz' inequality we get 
\begin{eqnarray*}  
|f(t,x)| &=&   
\left|\intop\limits_{\phantom{.}x_0}^{x}  
\frac\partial{\partial\xi} f(t,\xi)  
\kern1mm {\rm d} \xi \right|			  
\kern2mm\le\kern2mm \intop\limits_0^{2\pi}\left|  
\frac\partial{\partial x} f(t,x)  
\right|\kern1mm {\rm d} x			\\  
&\le& \left(\intop\limits_0^{2\pi}1^2  
\kern1mm {\rm d} x\right)^{\frac12}\cdot  
\left(  
\intop\limits_0^{2\pi}\left|  
\frac\partial{\partial x} f(t,x)  
\right|^2\kern1mm {\rm d} x  
\right)^{\frac12}  
\end{eqnarray*}  
that is 
\begin{eqnarray*}  
|f(t,x)|^2 &\le& 2\pi\intop\limits_0^{2\pi}\left|  
\frac\partial{\partial x} f(t,x)  
\right|^2\kern1mm {\rm d} x.  
\end{eqnarray*}  
\end{beweis}  
 
Now we are able to show boundedness of~$\psi$, 
uniformly on every strictly positive intervall. The key element 
in the proof is the investigation of a certain energy 
functional corresponding to~(\ref{Wellenpsi}). 
  
\begin{theorem}  
To every function~$\psi\in\mathcal C^2(\mathbb R_+ 
\times\mathbb T^1;\mathbb R)$, satisfying both~(\ref{mittellos})  
and~(\ref{Wellenpsi}), there is a positive constant~$C_{t_0}$ 
depending only on~$t_0>0$, such that for every~$t\ge t_0$ and  
all $x\in\mathbb T^1$ the inequality 
\label{Satz4} 
\begin{eqnarray}  
|\psi(t, x)|<C_{t_0} \label{Schaetz} 
\end{eqnarray}  
holds.
\end{theorem} 
  
\begin{remark} 
One verifies that equation (\ref{Wellenpsi}) has unbounded 
solutions, too. But none of them satisfies~(\ref{mittellos}) 
which underlines the importance of this assumption.
\end{remark}  
 
\begin{beweis} 
Let us define the energy 
\begin{eqnarray*} 
\varepsilon(t)&\mathop{=}\limits_{\rm Df}&\frac12  
\int_0^{2\pi}\left({\psi_t}^2+{\psi_x}^2\right){\rm d}x. 
\end{eqnarray*}   
Our smoothness assumptions in~$\psi$ guarantees 
differentiability of~$\varepsilon$ in $\mathbb R_+$.  
Integrating by parts and using equation~(\ref{Wellenpsi}) yields 
\begin{eqnarray*}  
\frac{\rm d}{{\rm d} t}\varepsilon&=& 
-\frac1{4 t^2}\int \limits_0^{2\pi} 
\psi_t \psi\kern1mm{\rm d}x. 
\end{eqnarray*}  
Since $\frac12({\psi_t}^2+\psi^2)+\psi_t\psi\ge0$ we have 
\begin{eqnarray*} 
\frac{\rm d}{{\rm d} t}\varepsilon  
&\le& 
\frac1{4 t^2}\cdot\frac12\int \limits_0^{2\pi} 
\left({\psi_t}^2+\psi^2\right)\kern1mm{\rm d}x.  
\end{eqnarray*}  
Now, lemma~\ref{Hilfssatz2} permits an estimate of the left 
hand side of the inequality in terms of $\varepsilon$ itself: 
\begin{eqnarray*} 
\frac{\rm d}{{\rm d} t}\varepsilon&\le&\frac{\pi^2}{t^2}\varepsilon  
\end{eqnarray*}  
Since $t\ge t_0>0$ we find  
\begin{eqnarray*}  
\varepsilon (t)&\le&  
\varepsilon (t_0)\kern1mm+\kern1mm  
\pi^2\intop\limits_{t_0}^t s^{-2}  
\varepsilon(s)\kern1mm {\rm d}s 
\end{eqnarray*}  
and Gronwall's inequality gives the uniform bound  
\begin{eqnarray*}  
\varepsilon (t)&\le&  
\varepsilon (t_0)\kern1mm+\kern1mm  
\varepsilon (t_0)\left({\rm e}^{\frac{\pi^2}{t_0}  
-\frac{\pi^2}{t\phantom{_0}}}  
-1\right)  
\kern2mm<\kern2mm  
\varepsilon (t_0)\kern1mm  
{\rm e}^{\frac{\pi^2}{t_0}}.  
\end{eqnarray*}  
Hence, the energy of an arbitrary solution is for every 
$t>t_0$ bounded and by definition non-negative.  
Now, up to a factor we can control the  
$L^2(\mathbb T^1)$-norm 
of~$\psi_x$ for all $t\ge t_0$ by the square root of the same 
constant and since~$\psi$ satisfies all suppositions of  
lemma~\ref{Hilfssatz2} a short calculation yield finally 
\begin{eqnarray*} 
|\psi(t,x)|&<&\sqrt{4\pi\kern.5mm\varepsilon(t_0)}\kern1mm  
{\rm e}^{\frac{\pi^2}{2t_0}}\kern2mm\equiv\kern2mm C_{t_0} 
\end{eqnarray*}  
for all $t\ge t_0>0$ and all $x\in\mathbb T^1$, hence $C_{t_0}$  
fulfills all needs. 
\end{beweis}  
 
We have shown that every spatially periodic solution of the 
Euler-Poisson-Darboux equation can be cast into the form 
\begin{eqnarray*} 
W(t,x)&=&\gamma\kern1mm+\kern1mm\beta\cdot\ln t 
\kern1mm+\kern1mmt^{-\frac12}\psi(t,x) 
\end{eqnarray*} 
where $\psi$ is bounded in $t\in[t_0,\infty)$. On the other 
hand for applications in the following sections we need deeper 
insights into the asymptotic behaviour of~$\psi$. 
Using the boundedness of~$\psi$ we can show that solutions 
to 
\begin{eqnarray*} 
\frac{\partial^2}{\partial t^2}\psi\kern1mm-\kern1mm 
\frac{\partial^2}{\partial x^2}\psi&=& 
-\frac1{4t^2}\psi 
\end{eqnarray*} 
behave like solutions of the free wave equation in two 
dimensions with a remainder falling off in time. 
  
\begin{theorem}  
Let $\psi\in\mathcal C^2(\mathbb R_+\times\mathbb T^1;\mathbb R)$  
be a bounded solution of \label{Satz5} 
\begin{eqnarray}  
\frac{\partial^2}{\partial t^2}\psi\kern1mm-\kern1mm 
\frac{\partial^2}{\partial x^2}\psi&=& 
-\frac1{4t^2}\psi,\label{Psi2} 
\end{eqnarray}  
with the side condition 
\begin{eqnarray*} 
\frac1{2\pi}\int_0^{2\pi}\psi(t,x)\kern1mm{\rm d}x&=&0 
\end{eqnarray*} 
for every $t\in\mathbb R_+$. Then there exist uniquely 
defined functions 
$\nu\in\mathcal C^2(\mathbb R_+\times\mathbb T^1;\mathbb R)$ and  
$\omega\in\mathcal C^2(\mathbb R_+\times\mathbb T^1;\mathbb R)$ 
as well as a constant~$C$ depending only on the choise of $t_0$  
such that  
\begin{eqnarray*}  
\psi(t,x)&=&\nu(t,x)\kern1mm+\kern1mm\omega(t,x)\\[2mm]  
\frac{\partial^2}{\partial t^2}\nu&=& 
\frac{\partial^2}{\partial x^2}\nu \\[2mm] 
|\omega(t,x)|&\le& C\cdot t^{-1}\\[2mm]  
|\omega_t (t,x)|&\le& C\cdot t^{-1}  
\end{eqnarray*} 
for all $t\in[t_0,\infty)$, $t_0>0$ and all $x\in\mathbb T^1$.  
\end{theorem} 
 
\begin{beweis}  
Let 
\begin{eqnarray*} 
u&\mathop{=}\limits_{\rm Df}&t\kern1mm+\kern1mm x\\[2mm] 
v&\mathop{=}\limits_{\rm Df}&t\kern1mm-\kern1mm x. 
\end{eqnarray*} 
In this coordinates~(\ref{Psi2}) reads 
\begin{eqnarray} 
\frac{\partial^2}{\partial u\kern1mm\partial v}\psi&=& 
-\frac14\frac{\psi}{(u+v)^2}.\label{Psi3} 
\end{eqnarray} 
As a consequence of this transformation some parallelogram,  
say~$\overline{1234}$ 
\ABB{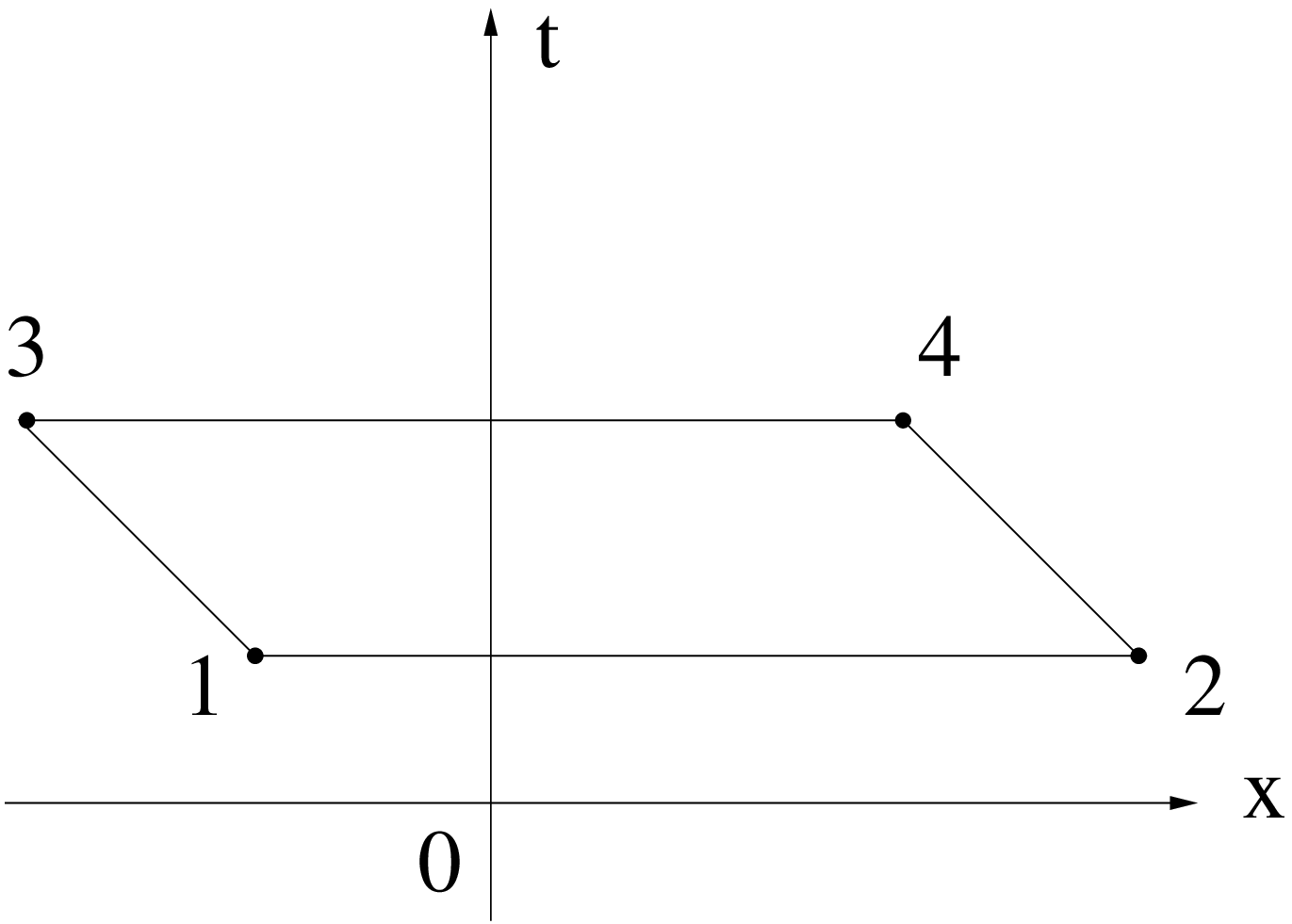}{10cm}  
{%Integrationsgebiet in\\ \phantom{eine kleine Verschiebung}  
%charakteristischen Koordinaten  
}  
\mbox{}\vskip-16mm  
\begin{eqnarray*}  
&&\kern45mm  
\begin{array}{ll}  
\left(t^{(1)}\equiv t_0,\right.&\left.x^{(1)}\equiv x_0=u_0-t_0  
\right)\\[2mm]  
\left(t^{(2)} = t_0,\right.&\left.x^{(2)}=x_0+2\pi  
\right)\\[2mm]  
\left(t^{(3)}= t_0+U,\right.&\left.x^{(3)}=x_0-U  
\right)\\[2mm]  
\left(t^{(4)}= t_0+U,\right.&\left.x^{(4)}=x_0+2\pi-U  
\right)  
\end{array}  
\end{eqnarray*}  
\vskip5mm  
 
has new vertices as follows: 
\ABB{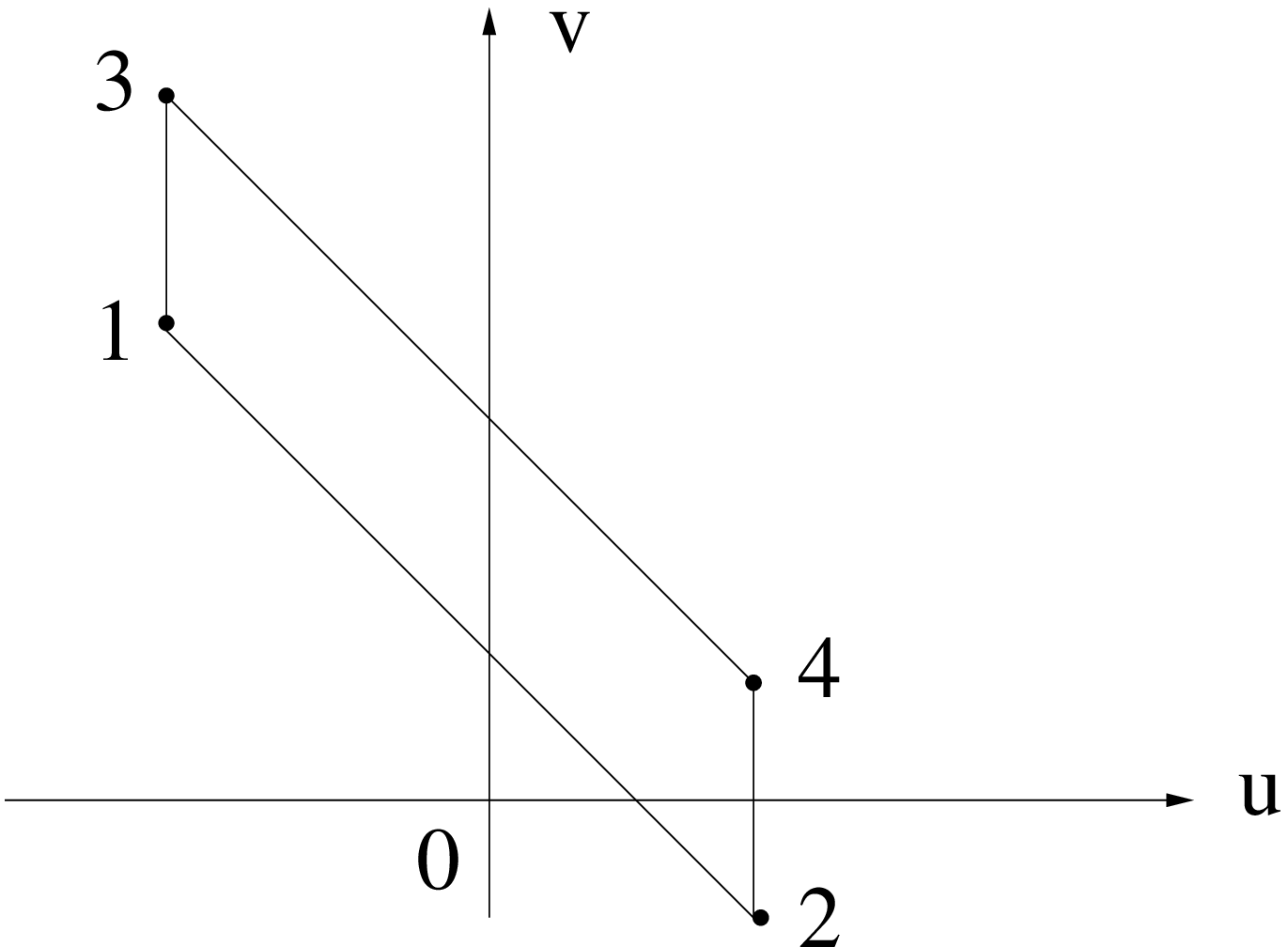}{10cm}  
{%Integrationsgebiet in\\ \phantom{eine kleine Verschiebung}  
%charakteristischen Koordinaten  
}  
\mbox{}\vskip-16mm  
\begin{eqnarray*}  
&&\kern45mm  
\begin{array}{ll}  
\left(u^{(1)}\equiv u_0,\right.&\left.v^{(1)}\equiv v_0=2t_0-u_0  
\right)\\[2mm]  
\left(u^{(2)} = u_0+2\pi,\right.&\left.v^{(2)}=v_0-2\pi  
\right)\\[2mm]  
\left(u^{(3)}= u_0,\right.&\left.v^{(3)}=v_0+2U  
\right)\\[2mm]  
\left(u^{(4)}= u_0+2\pi,\right.&\left.v^{(4)}=v_0-2\pi+2U  
\right)  
\end{array}  
\end{eqnarray*}  
\vskip7mm  
 
The periodicity in $x$-space is the identity 
\begin{eqnarray} 
(u,v)&\equiv&(u+2\pi n,v-2\pi n)\label{Periode} 
\end{eqnarray} 
for all $n\in\mathbb Z$ and the over-all condition 
$t>0$ converts to $u+v>0$. Both are generally assumed in what 
follows.~--~The estimate 
\begin{eqnarray*} 
|\psi_u(u,v_2)\kern1mm-\kern1mm 
\psi_u(u,v_1)|&\le&\intop_{v_1}^{v_2}\frac{|\psi|\kern1mm{\rm d}v} 
{(u+v)^2}\kern2mm\mathop{\le}\limits^{(\ref{Schaetz})}\kern2mm 
C_{t_0}\left(\frac1{u+v_1}\kern1mm-\kern1mm\frac1{u+v_2}\right) 
\end{eqnarray*} 
yields boundedness of $|\psi_u(u,v)|$ for any $u$ while 
$v\to\infty$. But the same inequality also proves with 
 $v_2=v_1+\Delta$, $\Delta>0$, the existence of a finite 
limit $F^\prime(u)$ for any $u$, hence 
\begin{eqnarray} 
F^\prime(u)&\mathop{=}\limits_{\rm Df}&   
\psi_u(u,2t_0-u)+\lim_{U\to\infty} 
\intop_{2t_0-u}^{2t_0-u+2U} \psi_{uv}(u,v) 
\kern1mm {\rm d}v \label{FStr} 
\end{eqnarray} 
is well-defined. Periodicity of $F^\prime(u)$ is due  
to~(\ref{Periode}) along with 
\begin{eqnarray*}  
F^\prime(u+2\pi)&=&   
\psi_u(u+2\pi,2t_0-u-2\pi)-\frac14  
\intop_{2t_0-u-2\pi}^\infty \frac{\psi(u+2\pi,v)}  
{(u+2\pi+v)^2}  
\kern1mm {\rm d}v\\[2mm] 
&=&   
\psi_u(u,2t_0-u)-\frac14  
\intop_{2t_0-u}^\infty \frac{\psi(u+2\pi,\tilde v-2\pi)}  
{(u+\tilde v)^2}  
\kern1mm {\rm d}\tilde v\\  
&=& F^\prime(u). 
\end{eqnarray*} 
Another property of $F^\prime(u)$ is its zero mean value. 
To show this we transform the defining integral back to 
$t$-$x$-coordinates according to the scetches from the beginning 
of this proof. Let $(u_0,v_0)$, that is $(t_0,x_0)$, 
$t_0>0$, an arbitrary point in characteristic or space-time 
coordinates, respectively. Then 
\begin{eqnarray}  
\intop_{u_0}^{u_0+2\pi} F^\prime(u)\kern1mm {\rm d}u  
&=& -\intop_{u_0}^{u_0+2\pi}   
\lim\limits_{U\to\infty}\intop\limits_{2t_0-u}^{2t_0-u+2U}  
\frac {\psi(u,v)}{4(u+v)^2}  
\kern1mm {\rm d}v\kern1mm {\rm d}u	\\  
&=&\lim\limits_{U\to\infty}\intop_{t_0}^{t_0+U}\frac1{2t^2}   
\intop\limits_{x_0+t_0-t}^{x_0+t_0-t+2\pi}  
\psi(t,x)  
\kern1mm {\rm d}x\kern1mm {\rm d}t\kern2mm=\kern2mm0, 
\label{F-Periode}  
\end{eqnarray} 
since the term under the integral is zero for every~$t$.  
A direct consequence is the periodicity of primitives. Let 
\begin{eqnarray*} 
\tilde F(u)&\mathop{=}\limits_{\rm Df}&\int F^\prime(u)\kern1mm 
{\rm d}u 
\end{eqnarray*} 
such an (arbitrary but fixed) primitive. Then the function 
\begin{eqnarray*} 
F(u)&\mathop{=}\limits_{\rm Df}&\tilde F(u) 
\kern1mm-\kern1mm\frac1{2\pi}\intop_{u_0}^{u_0+2\pi}  
\tilde F(u)\kern1mm{\rm d}u 
\end{eqnarray*} 
is defined uniquely and especially 
\begin{eqnarray} 
\intop_{u_0}^{u_0+2\pi} F(u)\kern1mm 
{\rm d}u&=&0 \label{FS-Periode} 
\end{eqnarray} 
is always satisfied. 
Now, starting with the inequality 
\begin{eqnarray*} 
|\psi_v(u_2,v)\kern1mm-\kern1mm 
\psi_v(u_1,v)|&\le&\intop_{u_1}^{u_2}\frac{|\psi|\kern1mm{\rm d}u} 
{(u+v)^2}\kern2mm\le\kern2mm 
C_{t_0}\left(\frac1{u_1+v}\kern1mm-\kern1mm\frac1{u_2+v}\right) 
\end{eqnarray*} 
we construct a likewise uniquely defined function 
$G(v)$, where all steps follow as above 
endowing $G$ with corresponding properties. Let us finally 
introduce the functions: 
\begin{eqnarray} 
\nu(u,v)&\mathop{=}\limits_{\rm Df}& 
F(u)\kern1mm+\kern1mm G(v)\label{Nue}\\[2mm] 
\omega(u,v)&\mathop{=}\limits_{\rm Df}& 
\psi(u,v)\kern1mm-\kern1mm \nu(u,v) 
\end{eqnarray} 
Obviously $\nu$ and $\omega$ are uniquely defined for any fixed 
$\psi$. $\nu$ solves the free wave equation $\nu_{uv}=0$ 
and regarding the remainder~$\omega$ we estimate for every $u$, $v$ 
subject to $u+v>0$ 
\begin{eqnarray}  
|\omega_u(u,v)|&=&|\psi_u(u,v)-\nu_u(u,v)|		 
\nonumber\\ [2mm] 
&=&\left|\psi_u(u,v) - \psi_u(u,2t_0-u) -  
\intop_{2t_0-u}^\infty\psi_{uv}(u,v)\kern1mm{\rm d}v\right|  
\nonumber\\[2mm]  
&=&\left|-\intop^{2t_0-u}_v  
\psi_{u\tilde v}(u,\tilde v)\kern1mm  
{\rm d}\tilde v\kern1mm-\kern1mm  
\intop_{2t_0-u}^\infty\psi_{uv}(u,v)\kern1mm{\rm d}v\right|	 
\nonumber\\ [2mm] 
&=&  
\left|\frac14\intop_v^\infty  
\frac{\psi(u,\tilde v)}{(u+\tilde v)^2}\kern1mm{\rm d}  
\tilde v\right|\kern2mm\le\kern2mm\frac{C_{t_0}}{4(u+v)},		  
\label{Itom} 
\end{eqnarray} 
where the estimate follows from~(\ref{Schaetz}). In exactly the same 
way one proves 
\begin{eqnarray*}  
|\omega_v(u,v)|&\le&\frac{C_{t_0}}{4(u+v)}.  
\end{eqnarray*}  
Now, since
\begin{eqnarray}  
|\omega_t(t,x)|&=& 
|\omega_u(u,v)+\omega_v(u,v)| 
\kern2mm\le\kern2mm\frac {C_{t_0}}{2(u+v)}\kern2mm=\kern2mm 
\frac{C_{t_0}}{4t}\label{Itom2}\\[2mm]  
|\omega_x(t,x)|&=& 
|\omega_u(u,v)-\omega_v(u,v)| 
\kern2mm\le\kern2mm\frac {C_{t_0}}{2(u+v)}\kern2mm=\kern2mm 
\frac{C_{t_0}}{4t}\label{Itom1} 
\end{eqnarray}  
for all $t\in[t_0,\infty)$, $x\in\mathbb T^1$ we have 
\begin{eqnarray*} 
||\omega_x(t,.)||_{L^2}^2&=& 
\intop_0^{2\pi} 
{\omega_x}^2\kern1mm{\rm d}x\kern2mm\le\kern2mm 
\frac{\pi {C_{t_0}}^2}{8t^2}. 
\end{eqnarray*} 
Corollary~\ref{om-Eigen} below proves the properties 
of~$\omega$ 
\begin{eqnarray*} 
\omega(t,x+2\pi)&=&\omega(t,x)	\\ 
\intop_0^{2\pi}\omega(t,x)\kern1mm{\rm d}x&=&0 
\end{eqnarray*} 
for every $(t,x)\in\mathbb R_+\times\mathbb T^1$. Hence all 
assumptions of lemma~\ref{Hilfssatz2} are satisfied which 
ensures the existence of some constant 
$C\mathop{=}\limits_{\rm Df}\frac\pi2 C_{t_0}>0$ with 
\begin{eqnarray} 
|\omega(t,x)|&\le& \frac Ct\label{A6} 
\end{eqnarray} 
for all $t\in[t_0,\infty)$, $t_0>0$ and all $x\in\mathbb T^1$. 
\end{beweis} 
 
The existence and uniqueness in the definition of function $\nu$ 
allow us to introduce the following term. 
 
\begin{definition}  
The function $\nu$ is called the  $\psi$-associated solution 
of the free wave equation, if~$\psi\in\mathcal C^2 
(\mathbb R_+\times\mathbb T^1;\mathbb R)$ solves 
\begin{eqnarray*} 
\frac{\partial^2}{\partial t^2}\psi\kern1mm-\kern1mm 
\frac{\partial^2}{\partial x^2}\psi&=& 
-\frac1{4 t^2}\psi 
\end{eqnarray*} 
with the side condition 
\begin{eqnarray*} 
\intop_0^{2\pi}\psi(t,x)\kern1mm{\rm d}x&=&0 
\end{eqnarray*} 
for every $t\in[t_0,\infty)$ and $\nu\in\mathcal C^2 
(\mathbb R_+\times\mathbb T^1;\mathbb R)$ is defined as in 
equation~(\ref{Nue}). 
\end{definition} 
 
Here follows a series of corollaries establishing properties 
of~$\nu$ as well as~$\omega$ which were used in the previous 
proof or are of frequent use in the sections below. 
 
\begin{corollary} 
Let $\nu$ be the $\psi$-associated solution of the free 
wave equation and $t_0>0$. Then there exists a positive constant 
$C$, such that for all $t\in[t_0,\infty)$ and $x\in\mathbb T^1$ 
the functions
$\nu$, $\nu_t$ and $\nu_x$ are bounded:\label{nu-Eigen} 
\begin{eqnarray} 
\mathop{\mathop{\rm max}\limits_{t\ge t_0>0}}\limits_{x\in\mathbb T^1} 
\kern1mm 
\left\{|\nu(t,x)|,|\nu_t(t,x)|,|\nu_x(t,x)|\right\}&<&C  \label{nuE1} 
\end{eqnarray} 
Further, for all $n\in\mathbb Z$ are: 
\begin{eqnarray} 
\nu(t,x+2\pi n)&=&\nu(t,x) 
\kern2mm=\kern2mm\nu(t+2\pi n,x)  \label{nuE2}	\\[2mm] 
\nu_t(t+2\pi n,x)&=&\nu_t(t,x)  \label{nuE4}	\\[2mm] 
\nu_x(t+2\pi n,x)&=&\nu_x(t,x)  \label{nuE5}	\\[2mm] 
\intop_0^{2\pi}\nu(t,x)\kern1mm{\rm d}x&=&0 \label{nuE3} 
\end{eqnarray} 
\end{corollary} 
 
\begin{beweis} 
 
(\ref{nuE1})\kern3mm 
By definition it is $\nu(u,v)=F(u)+G(v)$. $F$ is periodic with vanishing 
mean value. W.l.o.g.\ is $F(u_0)=0$, 
$\psi_u(u,v_\infty)\kern1mm\mathop{=}\limits_{\rm Df}\kern1mm 
\mathop{\rm lim}\limits_{v\to\infty}\psi_u(u,v)$. Then 
\begin{eqnarray*} 
F(u)&=&\intop_{u_0}^{u}F^\prime(\tilde u)\kern1mm{\rm d}\tilde u 
\kern2mm=\kern2mm 
\intop_{u_0}^{u}\psi_{\tilde u}(\tilde u,v_\infty) 
\kern1mm{\rm d}\tilde u 
\kern2mm=\kern2mm\psi(u,v_\infty)\kern1mm-\kern1mm\psi(u_0,v_\infty). 
\end{eqnarray*} 
Theorem~\ref{Satz4} shows boundedness of $\psi$  
(uniformly in the considered domain). 
The same is true for $G$. Finally, boundedness of $|\nu_t(t,x)|$ 
and $|\nu_x(t,x)|$ is a consequence of  
$|F^\prime(u)|$ and $|G^\prime(v)|$ boundedness established 
during the proof of theorem~\ref{Satz5}. 
 
(\ref{nuE2})\kern3mm 
With~(\ref{FS-Periode}) we have 
\begin{eqnarray*} 
\nu(t+2\pi n,x)&=&F(t+2\pi n+x)\kern1mm+\kern1mm G(t+2\pi n-x)\\ 
&=&F(t+x)\kern1mm+\kern1mmG(t-x) 
\kern2mm=\kern2mm\nu(t,x) 
\end{eqnarray*} 
and analogously 
\begin{eqnarray*} 
\nu(t,x+2\pi n)&=&F(t+x+2\pi n)\kern1mm+\kern1mm G(t-x-2\pi n)\\ 
&=&\nu(t,x). 
\end{eqnarray*} 
 
(\ref{nuE4})\kern3mm 
Is a direct consequence of (\ref{nuE2}). 
 
(\ref{nuE5})\kern3mm  
Follows directly from~(\ref{nuE2}). 
 
(\ref{nuE3})\kern3mm  
Property~(\ref{FS-Periode}) and correspondingly for~$G$ yield: 
\begin{eqnarray*} 
\intop_0^{2\pi} 
\nu(t,x)\kern1mm{\rm d}x&=& 
\intop_t^{t+2\pi} 
\nu(u,2t-u)\kern1mm{\rm d}u\\ 
&=& 
\intop_t^{t+2\pi}\left( 
F(u)\kern1mm+\kern1mm G(2t-u)\right) 
\kern1mm{\rm d}u\\ 
&=& 
\intop_t^{t+2\pi} 
F(u)\kern1mm{\rm d}u 
\kern1mm+\kern1mm 
\intop_{t-2\pi}^{t} G(v) 
\kern1mm{\rm d}v\kern2mm=\kern2mm0 
\end{eqnarray*} 
\end{beweis} 
 
\begin{corollary} 
Let $\nu$ be the $\psi$-associated solution of the free wave 
equation. Then the following are equivalent:\label{nu-Aequ} 
\begin{eqnarray} 
\psi(t,x)&\equiv&0    \label{nuA1}	\\[2mm] 
F(t+x)&\equiv&0 \kern5mm\text{and}\kern5mm 
G(t-x)\kern2mm\equiv\kern2mm0   \label{nuA2}	\\[2mm] 
\nu(t,x)&\equiv&0    \label{nuA3} 
\end{eqnarray} 
\end{corollary} 
 
\begin{beweis} 
 
(\ref{nuA1}) $\Rightarrow$ (\ref{nuA2}):\kern3mm  
$F$ is constant due to~(\ref{FStr}), 
from~(\ref{FS-Periode}) follows the statement with respect to~$F$;  
analogously for $G$. 
 
(\ref{nuA2}) $\Leftrightarrow$ (\ref{nuA3}):\kern3mm  
($\Rightarrow$) is due  
to~(\ref{Nue}) and ($\Leftarrow$)~due to~(\ref{Nue}) 
and~(\ref{FS-Periode}). 
 
(\ref{nuA3}) $\Rightarrow$ (\ref{nuA1}):\kern3mm 
It is $\psi(t,x)=\omega(t,x)$. The estimate~(\ref{Itom1}) 
reads here in a first step 
\begin{eqnarray*} 
|\psi_x(t,x)|_1&\le&\frac{C_{t_0}}{4t} 
\end{eqnarray*} 
and according to lemma~\ref{Hilfssatz2} 
\begin{eqnarray*} 
|\psi(t,x)|_1&\le&\frac{\pi C_{t_0}}{2t}. 
\end{eqnarray*} 
Now, this improved estimate can be used iteratively in~(\ref{Itom}), 
leading to an $n$-th order estimate 
\begin{eqnarray*} 
|\psi(t,x)|_n&\le&\left(\frac\pi 2\right)^n \frac{C_{t_0}}{n!} 
t^{-n} 
\end{eqnarray*} 
for every $x\in\mathbb T^1$ and every $t\in[t_0,\infty)$. 
Due to~(\ref{Itom2}) the same estimate is valid for $\psi_t$.  
Consequently exists to any 
$\varepsilon>0$ an $n_0$, such that 
\begin{eqnarray*} 
|\psi(t_0,x)|_n&<&\varepsilon\\ 
|\psi_t(t_0,x)|_n&<&\varepsilon 
\end{eqnarray*} 
for all $x\in\mathbb T^1$ and all $n\ge n_0$, that is 
$\psi(t_0,x)\kern1mm=\kern1mm\psi_t(t_0,x)\kern1mm=\kern1mm0$. 
Due to the uniqueness of its solutions 
$\psi$ is the trivial solution of~(\ref{Psi2}). 
\end{beweis} 
 
\begin{corollary} 
Let $\nu$ be the $\psi$-associated solution of the free wave 
equation, $\omega\mathop{=}\limits_{\rm Df}\psi-\nu$  
and $t_0>0$. Then there exists a positive constant 
$C$, such that for all $t\in[t_0,\infty)$, $x\in\mathbb T^1$ 
and $n\in\mathbb Z$:\label{om-Eigen} 
\begin{eqnarray} 
|\omega(t,x)|&<&C  \label{omE1}	\\[2mm] 
\omega(t,x+2\pi n)&=&\omega(t,x)  \label{omE2}	\\[2mm] 
\intop_0^{2\pi}\omega(t,x)\kern1mm{\rm d}x 
&=&0  \label{omE3} 
\end{eqnarray} 
\end{corollary} 
 
\begin{beweis} 
All properties are direct consequences from the definition  
of~$\omega$, the corresponding properties of~$\psi$ along 
with corollary~\ref{nu-Eigen} above. 
\end{beweis} 
 
\begin{corollary} 
Let $\nu$ be the $\psi$-associated solution of the free wave 
equation and $\omega\mathop{=}\limits_{\rm Df}\psi-\nu$.  
Then the following is equivalent:\label{om-Aequ} 
\begin{eqnarray} 
\psi(t,x)&\equiv&0   \label{omA1}	\\[2mm] 
\omega(t,x)&\equiv&0 \label{omA2} 
\end{eqnarray} 
\end{corollary} 
 
\begin{beweis} 
 
(\ref{omA1}) $\Rightarrow$ (\ref{omA2}):\kern3mm  
With corollary~\ref{nu-Aequ} is $\nu\equiv 0$. 
 
(\ref{omA2}) $\Rightarrow$ (\ref{omA1}):\kern3mm  
It is $\psi\equiv\nu$. Since 
\begin{eqnarray*} 
\frac{\partial^2}{\partial t^2}\psi\kern1mm-\kern1mm 
\frac{\partial^2}{\partial x^2}\psi&=&-\frac1{4t^2}\psi 
\kern3mm\mathop{\equiv}\limits^{!}\kern3mm0 
\end{eqnarray*} 
the statement follows. 
\end{beweis} 
 
Now we summarize the most important features in the 
asymptotic behaviour of the metric function~$W$. 
Here we use as remainder  
$\varkappa\mathop{=}\limits_{\rm Df}t^{-\frac12}\omega$. 
  
\begin{corollary}\label{Folgerung6}%  
Let $W\in\mathcal C^2(\mathbb R_+\times \mathbb T^1)$  
be a real-valued solution of the Euler-Poisson-Darboux 
equation  
\begin{eqnarray*}  
\frac{\partial^2}{\partial t^2}W(t,x)\kern1mm+\kern1mm  
\frac1t  
\frac{\partial}{\partial t}W(t,x)  
\kern1mm-\kern1mm  
\frac{\partial^2}{\partial x^2}W(t,x)&=&0.  
\end{eqnarray*}  
Then there exists uniquely defined, bounded, real-valued 
functions $\nu\in\mathcal C^2(\mathbb R_+\times\mathbb T^1)$  
and $\varkappa\in\mathcal C^2(\mathbb R_+\times\mathbb T^1)$,  
constants $\beta$, $\gamma$, and a positive constant $C_{t_0}$ 
such that 
\begin{eqnarray}  
W(t,x)&=&\gamma\kern1mm+\kern1mm\beta\cdot\ln t  
\kern1mm+\kern1mm t^{-\frac12}\nu(t,x)\kern1mm+\kern1mm  
\varkappa(t,x)\label{A1}\\  
\frac{\partial^2}{\partial t^2}\nu(t,x)&=&  
\frac{\partial^2}{\partial x^2}\nu(t,x)\label{A2}  
\end{eqnarray}  
and  
\begin{eqnarray}  
|\varkappa(t,x)|&\le&C_{t_0}\cdot t^{-\frac32}\label{A3}\\  
|\varkappa_t(t,x)|&\le&C_{t_0}\cdot t^{-\frac32}\label{A4}\\  
|\varkappa_x(t,x)|&\le&C_{t_0}\cdot t^{-\frac32}\label{A5}
\end{eqnarray}  
for all $t\ge t_0>0$ and all $x\in\mathbb T^1$.  
\end{corollary}  
\begin{proof}
We have already shown (\ref{A1}) and (\ref{A2}), explicitly. But on the
other hand estimates (\ref{A3}), (\ref{A4}) and (\ref{A5}) follow directly
from the above definition of~$\varkappa$ together with (\ref{A6}),
(\ref{Itom2}) and (\ref{Itom1}), respectively.
\end{proof}
 
\section{On the integration of constraints} 
 
As mentioned in the beginning part of this paper a special 
feature of Gowdy's model is the decoupling of the dynamical 
quantity~$a$. Furthermore, it is well-known that the 
momentum constraint 
\begin{eqnarray}   
\frac\partial{\partial x}a&=&\frac12\kern.5mm t \kern.5mm 
\frac\partial{\partial t}W\kern.5mm 
\frac\partial{\partial x}W \label{Impuls} 
\end{eqnarray}  
is conserved along $t$ developments; that means it is satisfied 
always if it is satisfied initially. 
 
With this remark in mind we will investigate now the    
Hamiltonian constraint 
\begin{eqnarray}  
\frac\partial{\partial t}a&=&-\frac1{4t}\kern1mm+  
\kern1mm\frac14\kern.5mmt 
\left[\left(\frac\partial{\partial t}W\right)^2 
\kern1mm+\kern1mm 
\left(\frac\partial{\partial x}W\right)^2\right] 
\label{Hamilton} 
\end{eqnarray}  
exclusively.  
  
\subsection*{Spatially homogeneous spacetimes} 
According to (\ref{Whom}) we find as the general solution 
of the Euler-Poisson-Darboux equation in spatially homogeneous 
spacetimes 
\begin{eqnarray*} 
W(t)&=&\beta\cdot\ln t\kern1mm+\kern1mm\gamma.  
\end{eqnarray*} 
The constraint (\ref{Hamilton}) reads 
\begin{eqnarray*}  
a_t(t)&=&\frac14(\beta^2-1)\cdot t^{-1},  
\end{eqnarray*}  
leading to 
\begin{eqnarray}  
a(t)&=&\frac14(\beta^2-1)\cdot\ln t\kern1mm+\kern1mm  
\zeta \label{ahom} 
\end{eqnarray}  
as solution where the arbitrary constant~$\zeta$ has to be 
specialized by an initial condition for $a$.  
  
\subsection*{Not spatially homogeneous spacetimes} 
The exceptional position of spatially homogeneous among 
the polarized Gowdy spacetimes is a consequence of the 
following consideration. 
 
\begin{theorem} 
Let $W\in\mathcal C^2(\mathbb R_+\times\mathbb T^1;\mathbb R)$ 
be a solution of the Euler-Poisson-Darboux equation. 
Then there is a positive constant~$C$, such that 
\begin{eqnarray*} 
|a_t(t,x)|&\le&C 
\end{eqnarray*} 
for all $t\in[t_0,\infty)$ and all $x\in\mathbb T^1$. 
\end{theorem} 
 
\begin{beweis} 
With corollary~\ref{Folgerung6} we have 
\begin{eqnarray*}  
W(t,x)&=&\gamma\kern1mm+\kern1mm\beta\cdot\ln t\kern1mm+\kern1mm  
t^{-\frac12}\nu(t,x)\kern1mm+\kern1mm\varkappa(t,x)  
\end{eqnarray*}  
with corresponding properties of $\nu$ and estimates for the
remainder~$\varkappa$; therefore
\begin{eqnarray}  
a_t&=&\frac14({\nu_t}^2+{\nu_x}^2)\kern1mm+\kern1mm\frac12  
\kern.5mm t^{-\frac12}\kern.5mm\beta\kern.5mm\nu_t     \nonumber\\ 
&&+\kern1mm\frac14 \kern.5mmt^{-1}\kern.5mm 
\left(\beta^2-1+2\nu_t\left(t^\frac32\varkappa_t-\frac12\nu\right) 
+2\nu_x t^\frac32\varkappa_x\right) \nonumber\\ 
&&+\kern1mm\mathcal O(t^{-\frac32}).                      \label{at}  
\end{eqnarray} 
The statement follows from corollaries~\ref{nu-Eigen}  
and~\ref{Folgerung6}. 
\end{beweis} 
 
\begin{remark} 
The above equation~(\ref{at}) already reflects the qualitatively 
different asymptotic behaviour in the metric function~$a$ depending on the 
spatial homogenity or not spatial homogenity of the underlying 
model. This is so because corollary~\ref{nu-Aequ} along with
corollary~\ref{om-Aequ}  
states that a polarized Gowdy spacetime is spatially 
homogeneous if and only if 
\begin{eqnarray*} 
\psi\kern1mm\equiv\kern1mm0&\leftrightarrow& 
\nu\kern1mm\equiv\kern1mm0\kern2mm\leftrightarrow\kern2mm 
\varkappa\kern1mm\equiv\kern1mm0. 
\end{eqnarray*} 
\end{remark} 
 
Since the asymptotics in spatially homogeneous models has been
completely investigated we will in what follows w.l.o.g.\ 
assume that   
$\nu_t\kern1mm=\kern1mm\nu_x\kern1mm=\kern1mm0$ is valid 
not everywhere. 
 
\begin{theorem} 
Let $a\in\mathcal C^2(\mathbb R_+\times\mathbb T^1;\mathbb R)$ 
be a solution of equation~(\ref{Hamilton}) and  
$\partial_x a\kern1mm=\kern1mm0$ not everywhere. Then there 
exist positive constants $\bar\nu$ and $C$ along with a uniquely 
determined function $\delta\in\mathcal C^2(\mathbb R_+\times 
\mathbb T^1;\mathbb R)$ such that 
\begin{eqnarray*} 
a(t,x)&=&\bar\nu\cdot t\kern1mm+\kern1mm\delta(t,x) 
\end{eqnarray*} 
with 
\begin{eqnarray*} 
|\delta(t,x)|&\le&C\cdot (1+t^\frac12) 
\end{eqnarray*} 
for every $t\in[t_0,\infty)$, $t_0>0$ and every $x\in\mathbb T^1$. 
\end{theorem} 
 
\begin{beweis} 
Since $a$ is continously differentiable, there is in  
$\mathbb R_+\times\mathbb T^1$ an open domain with 
$a_x\kern1mm\neq\kern1mm0$ hence $\nu_x\kern1mm\neq\kern1mm0$. 
Consequently 
\begin{eqnarray}  
\bar\nu&\mathop{=}\limits_{\rm Df}&\frac1{8\pi}  
\intop_{t_0}^{t_0+2\pi}({\nu_t}^2+{\nu_x}^2)\kern1mm{\rm d}t  
\label{nuq} 
\end{eqnarray} 
is a positive constant since it is also $x$-independent due to 
$\nu_{tt}\kern1mm=\kern1mm\nu_{xx}$ and 
\begin{eqnarray*}  
\frac{\rm d}{{\rm d}x}\bar\nu&=&\frac1{4\pi}  
\int_{t_0}^{t_0+2\pi}(\nu_t\nu_{tx}+\nu_x\nu_{xx})\kern1mm  
{\rm d}t\\  
&=&\frac1{4\pi}\int_{t_0}^{t_0+2\pi}\nu_x(-\nu_{tt}+\nu_{xx})  
\kern1mm{\rm d}t\kern1mm+\kern1mm\left.\frac1{4\pi}\kern.5mm\nu_t  
\kern.5mm\nu_x\kern.5mm\right|_{t_0}^{t_0+2\pi}\kern2mm=\kern2mm0. 
\end{eqnarray*}  
Let be further 
\begin{eqnarray*}  
\delta_t(t,x)&\mathop{=}\limits_{\rm Df}& 
-\bar\nu\kern1mm+\kern1mm a_t(t,x)\\ 
&=&-\bar\nu\kern1mm+\kern1mm\frac14({\nu_t}^2+{\nu_x}^2)\\ 
&&+\kern1mm\frac12  
\kern.5mm t^{-\frac12}\kern.5mm\beta\kern.5mm\nu_t \\ 
&&+\kern1mm\frac14 \kern.5mmt^{-1}\kern.5mm 
\left(\beta^2-1\kern.5mm+\kern.5mm 
2\nu_t\left(t^\frac32\varkappa_t-\frac12\nu\right) 
\kern.5mm+\kern.5mm2\nu_x t^\frac32\varkappa_x\right) \\ 
&&+\kern1mm\frac12\kern.5mmt^{-\frac32}\kern.5mm\beta\kern.5mm 
\left(t^\frac32\varkappa_t\kern.5mm-\kern.5mm\frac12\nu\right) \\ 
&&+\kern1mm\frac14\kern.5mmt^{-2}\left[\left(t^\frac32\varkappa_t- 
\frac12\nu\right)^2\kern.5mm+\kern.5mm\left(t^\frac32\varkappa_x 
\right)^2\right] 
\end{eqnarray*} 
which is uniquely determined for every fixed solution $W$  
of the Euler-Poisson-Darboux equation. So we have 
\begin{eqnarray*} 
\delta(t,x)&=&\delta(t_0,x)\kern1mm+\kern1mm\frac14\int_{t_0}^t 
\left({\nu_\tau}^2\kern.5mm+\kern.5mm{\nu_x}^2\right) 
\kern1mm{\rm d}\tau\kern1mm-\kern1mm\bar\nu\cdot(t-t_0) 
\kern1mm+\kern1mm\mathcal O(t^\frac12). 
\end{eqnarray*} 
Now we consider the second and third term of the 
right hand side using~(\ref{nuq}) in detail: 
\begin{eqnarray*}  
\lefteqn{\mbox{}\kern-12mm\frac14\int_{t_0}^t\left({\nu_\tau}^2 
\kern.5mm+\kern.5mm 
{\nu_x}^2\right)\kern1mm{\rm d}\tau\kern1mm-\kern1mm 
\bar\nu\cdot(t-t_0)}		\\[5mm] 
\mbox{}\kern12mm&=& 
\int_{t_0}^t\left[\frac14({\nu_\tau}^2+{\nu_x}^2)-\frac1{8\pi}  
\intop_{t_0}^{t_0+2\pi}({\nu_t}^2+{\nu_x}^2)\kern1mm{\rm d}t  
\right]\kern1mm{\rm d}\tau    \\[5mm] 
\mbox{}\kern12mm&=& 
\frac14\int_{t_0}^{t_0+2\pi\left[\frac{t-t_0}{2\pi}\right]}  
({\nu_t}^2+{\nu_x}^2)\kern1mm  
{\rm d}t\kern1mm-\kern1mm\frac{\left[\frac{t-t_0}{2\pi}\right]}{4}  
\intop_{t_0}^{t_0+2\pi}({\nu_t}^2+{\nu_x}^2)\kern1mm{\rm d}t\\[2mm] 
\mbox{}\kern12mm&&+  
\int_{t_0+2\pi\left[\frac{t-t_0}{2\pi}\right]}^t 
\left[\frac14({\nu_\tau}^2+{\nu_x}^2)  
\kern1mm-\kern1mm\frac1{8\pi}  
\intop_{t_0}^{t_0+2\pi}({\nu_t}^2+{\nu_x}^2)\kern1mm{\rm d}t  
\right]\kern1mm{\rm d}\tau  
\end{eqnarray*}  
Here as usual $[n]$ is the largest integer not 
greater then $n$. Now, due to periodicity the first two terms cancel 
each other. The integrand of the third is bounded, the domain 
of integration less then $2\pi$ so it is  
\begin{eqnarray*}  
\left|\intop_{t_0}^t\left[\frac14({\nu_\tau}^2+{\nu_x}^2)  
\kern1mm-\kern1mm\bar\nu  
\right]\kern1mm{\rm d}\tau\right|&\le&2\pi C 
\end{eqnarray*}  
and finally 
\begin{eqnarray*}  
|\delta(t,x)|&\le&C\cdot(1+t^\frac12)  
\end{eqnarray*}  
for a positive constant $C$, all $x\in\mathbb T^1$  
and every~$t\in[t_0,\infty)$, $t_0>0$. 
\end{beweis} 
 
All results concerning the integration of constrains are 
summarized by the following theorem.  
  
\begin{theorem}\label{Satz7}%  
Let $W\in\mathcal C^2(\mathbb R_+\times\mathbb T^1;\mathbb R)$  
be a solution of $W_{tt}+t^{-1}W_t-W_{xx}=0$.  
Then any solution   
$a\in\mathcal C^2(\mathbb R_+\times\mathbb T^1;\mathbb R)$ 
of  
\begin{eqnarray*}  
a_t&=&-\frac14t^{-1}\kern1mm+\kern1mm\frac14t\kern.5mm  
({W_t}^2+{W_x}^2)  
\end{eqnarray*}  
may be cast for all $t\ge t_0>0$ and all $x\in\mathbb T^1$  
into the form 
\begin{eqnarray}  
a(t,x)&=&\left\{  
\begin{array}{ll}  
\frac14(\beta^2-1)\cdot\ln t\kern1mm+\kern1mm\zeta,&  
\text{if }\partial_x a(t,x)\equiv0\\[2mm]  
\bar\nu\cdot t\kern1mm+\kern1mm\delta(t,x),&\text{otherwise}  
\end{array}  
\right. \label{aasym} 
\end{eqnarray}  
where $\bar\nu$ is a positive, $\zeta$ an arbitrary constant 
and $\delta$ satisfies the inequalities 
\begin{eqnarray*}  
|\delta(t,x)|&\le&C\cdot (1+t^\frac12)\\[2mm]  
|\delta_t(t,x)|&\le&C  
\end{eqnarray*}  
with another positive constante~$C$.  
\end{theorem} 
  
\section{On geodesic completeness in polarized Gowdy spacetimes}  
 
In contrast to the analytical treatments in the first part 
of this paper we will now investigate some open 
geometrical questions in polarized Gowdy spacetimes. 
 
Isenberg and Moncrief~(1990) established important results 
with respect to the singular behaviour in this class of  
spacetimes. However, the question concerning possible 
singularities in its future development remained open 
(at least for the case with spatial $\mathbb T^3$-topology) and 
shall be the subject of this section. As usual we will 
identify the term ``singularity'' with the term 
``existence of an uncomplete non-spacelike geodesic'' 
and use the basic concepts and relations of the theory 
of Lorentzian manifolds and causality without repeating 
all its definitions. However, for convenience we will summarize  
some facts for short.  
 
Future will always mean the expanding direction 
of the universe. 
 
In local coordinates 
$(x^0,x^1,x^2,x^3)$ the geodesic equation reads: 
\begin{eqnarray} 
\frac{{\rm d}^2 x^i}{{\rm d}\tau^2}\kern1mm+\kern1mm 
\sum_{j,k}\kern.5mm\Gamma^i_{jk}\kern.5mm 
\frac{{\rm d} x^j}{{\rm d}\tau}\frac{{\rm d} x^k}{{\rm d}\tau} 
&=&0,\kern10mmi,j,k=0,\dots,3\label{Geo} 
\end{eqnarray} 
Among the properties of affinly parametrized geodesics $\mathfrak x$
the following are of special importance: 
 
1.) The tangent vector is normalized 
\begin{eqnarray} 
\sum_{i,j}\kern.5mmg_{ij}\kern.5mm\frac{{\rm d}x^i}{{\rm d}\tau} 
\kern.5mm\frac{{\rm d}x^j}{{\rm d}\tau}&=&-\varepsilon, 
\label{Norm} 
\end{eqnarray} 
where $\varepsilon=0$, if the geodesic is null and 
$\varepsilon=1$, if it is timelike. 
 
2.) The scalar product of any Killing vectorfield $X$ and the 
geodesic tangent is constant along the curve; in coordinates: 
\begin{eqnarray} 
\frac{{\rm d}}{{\rm d}\tau}\kern.5mm\sum_{i,j}\kern.5mmg_{ij} 
\kern.5mm{X_\mathfrak x}^i\kern.5mm\frac{{\rm d}x^j}{{\rm d}\tau} 
&=&0 
\label{KVF} 
\end{eqnarray} 
 
Gowdy spacetimes are known to be maximally extended, globally  
hyperbolic regions which permit foliations by  
$(t={\rm const})$-hypersurfaces. Consequently, any nonspacelike 
curve intersects every such hypersurface exactly once. 
This is important in view of the possibility to take 
$t\to\infty$ limits along an arbitrary nonspacelike curve. 
Furthermore, according to our choise of orientation it is  
\begin{eqnarray} 
\frac{{\rm d}t}{{\rm d}\tau}&>&0 \label{Orient} 
\end{eqnarray} 
if $\tau$ is as before the parameter of the future directed curve. 
 
Now we can start with the investigations. Since the asymptotic 
expansion of polarized Gowdy spacetimes is quite different depending 
on the spatial homogenity/inhomogenity of the model it is 
necessary to split the proofs accordingly. 
 
\subsection*{Geodesics in spatially homogeneous, polarized 
spacetimes}  
Spatially homogeneous, polarized spacetimes correspond to 
exactly solvable Einstein equations. We had found~(\ref{Whom})  
and~(\ref{ahom}), 
\begin{eqnarray}  
W(t)&=&\gamma\kern1mm+\kern1mm\beta\cdot\ln t\label{Whom1}\\  
a(t)&=&\zeta\kern1mm+\kern1mm\frac14(\beta^2-1)\cdot\ln t,  
\label{ahom1} 
\end{eqnarray}  
with constants $\beta$, $\gamma$, $\zeta$. We need the 
following estimate to prove geodesic completeness in such 
models.  
  
\begin{lemma}  
Let $\left(\mathbb R_+\times\mathbb T^3,g={\rm diag}(g_{00},g_{11},g_{22},g_{33})\right)$  
be a spatially homogeneous solution of Einstein's field equation  
for the polarized Gowdy model. Then there exist a constant~$C_\beta$, 
such that the Christoffel symbols satisfy for every 
$t\in[t_0,\infty)$:\label{Voll1} 
\begin{eqnarray*}   
\Gamma^0_{11}(t)&>&-\frac12 \kern1mm g^{00}\kern1mm 
C_\beta\kern1mm t^{-1}\cdot g_{11} \\ [2mm] 
\Gamma^0_{22}(t)&>&-\frac12 \kern1mm g^{00}\kern1mm  
C_\beta\kern1mm t^{-1}\cdot g_{22} \\ [2mm] 
\Gamma^0_{33}(t)&>& -\frac12\kern1mm g^{00}\kern1mm 
C_\beta\kern1mm t^{-1}\cdot g_{33} 
\end{eqnarray*}  
\end{lemma}  
  
\begin{beweis}  
The following section~\ref{Symbole} contains the relevant 
Christoffel symbols. 
Using~(\ref{Whom1}) and~(\ref{ahom1}) we find: 
\begin{eqnarray*}  
\Gamma^0_{11}(t)&=&-\frac12g^{00}\partial_t g_{11} 
\kern2mm=\kern2mm-\frac12g^{00}\cdot\frac12(\beta^2-1)\kern1mm  
t^{-1}\cdot g_{11}\\[2mm]   
\Gamma^0_{22}(t)&=&-\frac12g^{00}\partial_t g_{22} 
\kern2mm=\kern2mm-\frac12g^{00}\cdot\phantom{\frac12^2} 
(\beta+1)\kern1mm 
t^{-1}\cdot g_{22}\\ [2mm]  
\Gamma^0_{33}(t)&=&-\frac12g^{00}\partial_t g_{33} 
\kern2mm=\kern2mm-\frac12g^{00}\cdot\phantom{\frac12} 
(-\beta+1)\kern1mm 
t^{-1}\cdot g_{33} 
\end{eqnarray*}   
The constant $C_\beta$, 
\begin{eqnarray} 
C_\beta&\mathop{=}\limits_{\rm Df}&-\frac12(\beta^2+3),\label{Cbeta} 
\end{eqnarray} 
proves the statement. Observe that~$C_\beta<0$. 
\end{beweis} 
  
\begin{theorem}  
Every inextendable, future directed, nonspacelike geodesic in an 
arbitrary but spatially homogeneous, polarized Gowdy manifold 
is in its future direction complete.  
\end{theorem}  
  
\begin{beweis}  
It is sufficient to regard the $t$-component of an arbitrary 
nonspacelike geodesic with affin parameter~$\tau$. Due to the 
assumed homogenity not all of the symbols in section~\ref{Symbole} 
are different from zero. More precisely, the $0$-component  
of~(\ref{Geo}) reads 
\begin{eqnarray*}   
\frac{{\rm d}^2t}{{\rm d}\tau^2}\kern1mm+\kern1mm   
\Gamma^0_{00}\left(\frac{{\rm d}t}{{\rm d}\tau}\right)^2\kern1mm+\kern1mm   
\Gamma^0_{11}\left(\frac{{\rm d}x}{{\rm d}\tau}\right)^2\kern1mm+\kern1mm   
\Gamma^0_{22}\left(\frac{{\rm d}y}{{\rm d}\tau}\right)^2\kern1mm+\kern1mm   
\Gamma^0_{33}\left(\frac{{\rm d}z}{{\rm d}\tau}\right)^2&=&0. 
\end{eqnarray*}   
The causal character of the curve and the choise of time orientation 
imply~(\ref{Orient}) and further 
\begin{eqnarray}  
\lefteqn{%\mbox{}\kern-4mm 
\frac{\rm d}{{\rm d}t}\ln\left(\frac{{\rm d}t}{{\rm d}\tau}\right)^{-1} 
\kern2mm=}\nonumber\\[2mm] 
&&\kern-12mm\left(\frac{{\rm d}t}{{\rm d}\tau}\right)^{-2}\left[  
\Gamma^0_{00}\left(\frac{{\rm d}t}{{\rm d}\tau}\right)^2\kern1mm+\kern1mm  
\Gamma^0_{11}\left(\frac{{\rm d}x}{{\rm d}\tau}\right)^2\kern1mm+\kern1mm  
\Gamma^0_{22}\left(\frac{{\rm d}y}{{\rm d}\tau}\right)^2\kern1mm+\kern1mm  
\Gamma^0_{33}\left(\frac{{\rm d}z}{{\rm d}\tau}\right)^2  
\right].\label{kern} 
\end{eqnarray}  
Now, using lemma~\ref{Voll1} we can estimate the symbols and get   
\begin{eqnarray*}   
\lefteqn{%  
\Gamma^0_{11}\left(\frac{{\rm d}x}{{\rm d}\tau}\right)^2\kern1mm+\kern1mm   
\Gamma^0_{22}\left(\frac{{\rm d}y}{{\rm d}\tau}\right)^2\kern1mm+\kern1mm   
\Gamma^0_{33}\left(\frac{{\rm d}z}{{\rm d}\tau}\right)^2\kern2mm>\kern2mm}\\ 
&&\kern14mm-\frac12 g^{00}  
C_\beta \kern1mmt^{-1}\left[  
g_{11}\left(\frac{{\rm d}x}{{\rm d}\tau}\right)^2\kern1mm+\kern1mm  
g_{22}\left(\frac{{\rm d}y}{{\rm d}\tau}\right)^2\kern1mm+\kern1mm   
g_{33}\left(\frac{{\rm d}z}{{\rm d}\tau}\right)^2  
\right]. 
\end{eqnarray*}  
On the other hand we have~(\ref{Norm}) 
\begin{eqnarray*}  
g_{11}\left(\frac{{\rm d}x}{{\rm d}\tau}\right)^2\kern1mm+\kern1mm   
g_{22}\left(\frac{{\rm d}y}{{\rm d}\tau}\right)^2\kern1mm+\kern1mm   
g_{33}\left(\frac{{\rm d}z}{{\rm d}\tau}\right)^2&=&  
-\varepsilon-g_{00}\left(\frac{{\rm d}t}{{\rm d}\tau}\right)^2 \\ 
&<&-g_{00}\left(\frac{{\rm d}t}{{\rm d}\tau}\right)^2  
\end{eqnarray*}  
where $\varepsilon\in\{0,1\}$ so we get finally 
($C_\beta<0$)  
\begin{eqnarray*}  
\Gamma^0_{11}\left(\frac{{\rm d}x}{{\rm d}\tau}\right)^2\kern1mm+\kern1mm   
\Gamma^0_{22}\left(\frac{{\rm d}y}{{\rm d}\tau}\right)^2\kern1mm+\kern1mm   
\Gamma^0_{33}\left(\frac{{\rm d}z}{{\rm d}\tau}\right)^2  
&>&\frac12 C_\beta \kern1mmt^{-1}  
\left(\frac{{\rm d}t}{{\rm d}\tau}\right)^2.  
\end{eqnarray*}   
Using~(\ref{kern}) as well as~$\Gamma^0_{00}= 
\frac14(\beta-1)\cdot t^{-1}$ and~(\ref{Cbeta}) 
\begin{eqnarray*}   
\frac{\rm d}{{\rm d}t}\ln\left(\frac{{\rm d}t}{{\rm d}\tau}\right)^{-1}  
&>&\frac12  
\left[   
\frac12(\beta^2-1)+C_\beta  
\right]\kern.5mm t^{-1}\kern2mm=\kern2mm-t^{-1}  
\end{eqnarray*}   
we get an suitable estimate. Integrating twice the result is 
\begin{eqnarray*}  
\tau(t)&>&\tau(t_0)\kern1mm+\kern1mmC\cdot\ln\frac{t}{t_0}  
\end{eqnarray*}  
for some positive $C$. Obviously, by taking $t\to\infty$ we have 
proved that $\tau\to\infty$ holds along any nonspacelike 
geodesic.  
\end{beweis}  
   
\subsection*{Geodesics in not spatially homogeneous,  
polarized spacetimes}  
  
In the case under consideration the components of Gowdy metric 
are not everywhere $x$-independent and consequently, the 
solutions of Einstein equations are not known explicitly. 
Due to limitation on asymptotic behaviour as well as the 
more complex form of geodesic equation proofs are more technical here.  
  
We begin by poving some estimates: 
  
\begin{lemma} 
Let $t_0>0$ and $W\colon\mathbb R_+\times\mathbb T^1\to\mathbb R$  
some smooth solution of $W_{tt}+t^{-1}W_t-W_{xx}=0$,  
$W_x\equiv\kern-3.0mm/ \kern1.5mm0$. Then there is a positive  
constant $C$, such that  for all $t>t_0$  
\label{Voll2} 
\begin{eqnarray*}  
\left|-\frac1{2t^2}+\frac12({W_t}^2-{W_x}^2)\right|&\le&C\cdot t^{-1}  
\end{eqnarray*}  
holds. 
\end{lemma}  
  
\begin{beweis} 
For $t$ sufficiently large we know according to
corollary~\ref{Folgerung6}  
$W$ is of the form 
\begin{eqnarray} 
W(t,x)&=&\gamma\kern1mm+\kern1mm\beta\cdot\ln t\kern1mm+\kern1mm 
t^{-\frac12}\left(\nu(t,x)\kern1mm+\kern1mm\omega(t,x)\right). 
\label{W3} 
\end{eqnarray}  
Theorem~\ref{Satz4} proved boundedness of $\nu+\omega$ 
while its partial derivatives are bounded by theorem~\ref{Satz5}, 
hence the lemma.  
\end{beweis}  
  
\begin{lemma}  
For any polarized Gowdy spacetime model there is a constant~$C_1$  
and a positive constant~$C$, such that  
\label{Voll3} 
\begin{eqnarray*}  
|g_{ab}|\kern1mm+\kern1mm|\partial_t g_{ab}| 
\kern1mm+\kern1mm|g^{ab}|\kern1mm+\kern1mm  
|\partial_t g^{ab}|&\le&C\cdot t^{C_1}  
\end{eqnarray*}  
where $a,b\in\{2,3\}$ and $t_0<t<\infty$.  
\end{lemma}  
  
\begin{beweis} 
(\ref{W3}) shows logarithmic growth of $W$ if $\beta\neq0$, 
consequently the metric coefficients $g_{ab}$, its inverses 
and derivatives can increase at most with power~$\beta+1$.  
So a constant~$C_1\mathop{=}\limits_{\rm Df}|\beta|+1$ 
will be sufficient for our needs. 
\end{beweis} 
  
\begin{theorem}  
Every inextendable, future directed, nonspacelike geodesic in an 
arbitrary but not spatially homogeneous, polarized Gowdy manifold 
is in its future direction complete.  
\end{theorem}

\begin{beweis}  
Let $\mathfrak x=(t,x,y,z)$ be the components of some 
arbitrary nonspacelike geodesic in local coordinates, 
having affin parameter~$\tau$ and taking values in an 
likewise arbitrary, polarized (not spatially homogeneous)  
Gowdy manifold. As before it will sufficient to restrict 
considerations on the $t$-component. Here it holds:  
\begin{eqnarray}  
0&=&\frac{{\rm d}^2t}{{\rm d}\tau^2}\kern1mm+\kern1mm\Gamma^0_{00}  
\left(\frac{{\rm d}t}{{\rm d}\tau}\right)^2\kern1mm+\kern1mm  
2\kern.5mm\Gamma^0_{01}\kern.5mm\frac{{\rm d}t}{{\rm d}\tau}  
\frac{{\rm d}x}{{\rm d}\tau}\kern1mm+\kern1mm  
\Gamma^0_{11}\left(\frac{{\rm d}x}{{\rm d}\tau}\right)^2\nonumber\\  
&&+\kern1mm  
\Gamma^0_{22}\left(\frac{{\rm d}y}{{\rm d}\tau}\right)^2\kern1mm+\kern1mm  
\Gamma^0_{33}\left(\frac{{\rm d}z}{{\rm d}\tau}\right)^2 
\label{kern2} 
\end{eqnarray}  
Again, Christoffel symbols are taken from section~\ref{Symbole}. 
Using constraints~(\ref{Hamilton}) and~(\ref{Impuls})  
we get: 
\begin{eqnarray*}  
\lefteqn{%  
\mbox{}\kern-25mm 
\Gamma^0_{00}\left(\frac{{\rm d}t}{{\rm d}\tau}\right)^2\kern1mm+\kern1mm  
2\kern.5mm\Gamma^0_{01}\kern.5mm\frac{{\rm d}t}{{\rm d}\tau}  
\frac{{\rm d}x}{{\rm d}\tau}\kern1mm+\kern1mm  
\Gamma^0_{11}\left(\frac{{\rm d}x}{{\rm d}\tau}\right)^2}\\[2mm]  
\mbox{}\kern25mm&=&    
\frac12 t\left(W_t\frac{{\rm d}t}{{\rm d}\tau}+   
W_x\frac{{\rm d}x}{{\rm d}\tau}\right)^2   
\kern1mm-\kern1mm\frac1{4t}  
\left[\left(\frac{{\rm d}t}{{\rm d}\tau}\right)^2+   
\left(\frac{{\rm d}x}{{\rm d}\tau}\right)^2\right]\\[2mm]  
&& -\kern1mm   
\frac t4({W_t}^2-{W_x}^2)   
\left[\left(\frac{{\rm d}t}{{\rm d}\tau}\right)^2-  
\left(\frac{{\rm d}x}{{\rm d}\tau}   
\right)^2\right]  
\end{eqnarray*} 
The first term on the right hand side is nonnegative and 
will be neglected. Using condition~(\ref{Norm}) 
in the form  
\begin{eqnarray*}  
\left(\frac{{\rm d}x}{{\rm d}\tau}\right)^2&=&  
\left(\frac{{\rm d}t}{{\rm d}\tau}\right)^2  
\kern1mm+\kern1mm g^{00} 
\left[\varepsilon\kern.5mm+\kern.5mm  
g_{22}\left(\frac{{\rm d}y}{{\rm d}\tau}\right)^2  
\kern.5mm+\kern.5mm 
g_{33}\left(\frac{{\rm d}z}{{\rm d}\tau}\right)^2\right]  
\end{eqnarray*}   
where as before $\varepsilon\in\{0,1\}$ depending on causal 
character of geodesic, we get  
\begin{eqnarray*}  
\Gamma^0_{00} \left(\frac{{\rm d}t}{{\rm d}\tau}\right)^2\kern1mm+\kern1mm  
2\kern.5mm\Gamma^0_{01}\kern.5mm\frac{{\rm d}t}{{\rm d}\tau}  
\frac{{\rm d}x}{{\rm d}\tau}\kern1mm+\kern1mm  
\Gamma^0_{11}\left(\frac{{\rm d}x}{{\rm d}\tau}\right)^2 
&\ge&-\frac1{2t}\kern.5mm  
\left(\frac{{\rm d}t}{{\rm d}\tau}\right)^2		\\ 
&&\mbox{}\kern-65mm+\left[-\frac1{4t}\kern.5mm+\kern.5mm  
\frac t4({W_t}^2-{W_x}^2)\right]  
\kern.5mm g^{00} \kern.5mm  
\left[\varepsilon\kern.5mm+\kern.5mm  
g_{22}\left(\frac{{\rm d}y}{{\rm d}\tau}\right)^2  
\kern.5mm+\kern.5mm  
g_{33}\left(\frac{{\rm d}z}{{\rm d}\tau}   
\right)^2\right].  
\end{eqnarray*}  
This estimate yield for the factor  
($2t^{-1}$) of geodesic equation~(\ref{kern2}) 
\begin{eqnarray}  
\lefteqn{% 
\frac{\rm d}{{\rm d}t}\left[t^{-1}\left(  
\frac{{\rm d}t}{{\rm d}\tau}\right)^2\right]  
\kern2mm\le}\nonumber\\[5mm] 
&&\mbox{}\kern10mm 
-g^{00}\left[-\frac1{2t^2}+\frac 12({W_t}^2-{W_x}^2)\right]  
\left[\varepsilon\kern.5mm+\kern.5mm  
g_{22}\left(\frac{{\rm d}y}{{\rm d}\tau}\right)^2\kern.5mm+\kern.5mm 
g_{33}\left(\frac{{\rm d}z}{{\rm d}\tau}\right)^2  
\right]\nonumber\\[2mm]  
&&\mbox{}\kern10mm 
+g^{00}\left[\partial_t  
g_{22}\cdot t^{-1}\kern.5mm 
\left(\frac{{\rm d}y}{{\rm d}\tau}\right)^2\kern.5mm+\kern.5mm  
\partial_t  
g_{33} 
\cdot t^{-1}\kern.5mm\left(\frac{{\rm d}z}{{\rm d}\tau}\right)^2   
\right],\label{Voll4} 
\end{eqnarray}  
where we have used~(\ref{Orient}). 
 
Due to~(\ref{KVF}) the scalar products~$g(\dot\mathfrak x,\cdot)$  
in 
\begin{eqnarray*} 
\left(\frac{{\rm d}y}{{\rm d}\tau}\right)^2  
&=&  
\left[g^{22}\cdot g\left(  
\dot\mathfrak x,\frac\partial{\partial y}\right)\right]^2\\[2mm]  
\left(\frac{{\rm d}z}{{\rm d}\tau}\right)^2   
&=&  
\left[g^{33}\cdot g\left(  
\dot\mathfrak x,\frac\partial{\partial z}\right)\right]^2  
\end{eqnarray*}  
are constant along the geodesic. 
Thanks to lemmas~\ref{Voll2} and~\ref{Voll3} we can estimate 
all brackets on the right hand side of the inequality~(\ref{Voll4})  
by an $\mathcal O(t^C)$-term, where the constant~$C$ only depends 
on the model parameter $\beta$ under consideration.  
Furthermore, due to~(\ref{aasym}) it holds 
$g^{00}\sim\mathcal O ({\rm e}^{-2\bar\nu t})$, $\bar\nu>0$, 
$g^{00}$ will dominate the development on the right hand side.
Consequently there are positive constants
$C_1$, $C_2$ with
\begin{eqnarray*}   
\frac{\rm d}{{\rm d}t}\left[t^{-1}\left(   
\frac{{\rm d}t}{{\rm d}\tau}\right)^2\right]   
&\le&  
C_1\kern.5mm{\rm e}^{-C_2t}  
\end{eqnarray*}  
and furthermore some positive constant 
${C_3}^{-2}$, such that for sufficiently large $t$ it holds  
\begin{eqnarray*}  
t^{-1}\left(  
\frac{{\rm d}t}{{\rm d}\tau}  
\right)^2&\le&{C_3}^{-2}  
\end{eqnarray*}  
and a final integration gives
\begin{eqnarray*}  
\tau(t)&\ge&\tau(t_1)\kern1mm+\kern1mm2{C_3}\left(\sqrt{t}-  
\sqrt{t_1}\right),  
\end{eqnarray*}  
which proves $\tau\to\infty$ for $t\to\infty$.  
\end{beweis}  
  
We summarize as follows:  
  
\begin{corollary}  
Every inextendable, future directed, nonspacelike geodesic in an 
arbitrary polarized Gowdy manifold 
is in its future direction complete. 
\end{corollary}

\section{Nonvanishing Christoffel symbols in polarized Gowdy spacetimes} 
\label{Symbole} 
 
In polarized Gowdy models the nonvanishing components of the metric
tensor are 
\begin{eqnarray*}  
g_{00}&=&-{\rm e}^{2a}\kern2cm g_{22}\kern2mm=\kern2mmt\kern.5mm{\rm e}^{W}\\  
g_{11}&=&\phantom{-}{\rm e}^{2a}\kern2cm g_{33}\kern2mm=\kern2mmt\kern.5mm{\rm e}^{-W}  
\end{eqnarray*}  
where $a=a(t,x)$, $W=W(t,x)$, so up to symmetry the only nonvanishing
symbols are  
\begin{eqnarray*}  
\begin{array}{lcrclclcrcl}  
\Gamma^0_{00}&=&\frac12g^{00}\partial_t g_{00}&=&a_t&\phantom{mmm}&  
\Gamma^0_{11}&=&-\frac12g^{00}\partial_t g_{11}&=&a_t\\[2mm] 
\Gamma^0_{01}&=&\frac12g^{00}\partial_x g_{00}&=&a_x&\phantom{mmm}&  
\Gamma^1_{01}&=&\frac12g^{11}\partial_t g_{11}&=&a_t\\[2mm] 
\Gamma^1_{00}&=&-\frac12g^{11}\partial_x g_{00}&=&a_x&\phantom{mmm}&  
\Gamma^1_{11}&=&\frac12g^{11}\partial_x g_{11}&=&a_x  
\end{array}  
\end{eqnarray*}  
\begin{eqnarray*} 
\begin{array}{lcrcl} 
\Gamma^0_{22}&=&-\frac12g^{00}\partial_t g_{22}&=&-\frac12g^{00}  
(t^{-1}+W_t)g_{22}\\[2mm] 
\Gamma^2_{02}&=&\frac12g^{22}\partial_t g_{22}&=&\phantom{-}\frac12(t^{-1}+W_t)\\[2mm]  
\Gamma^1_{22}&=&-\frac12g^{11}\partial_x g_{22}&=&-\frac12g^{11}  
W_x g_{22}\\[2mm]  
\Gamma^2_{12}&=&\frac12g^{22}\partial_x g_{22}&=&\phantom{-}\frac12W_x\\[2mm]  
\Gamma^0_{33}&=&-\frac12g^{00}\partial_t g_{33}&=&-\frac12g^{00}  
(t^{-1}-W_t)g_{33}\\[2mm]  
\Gamma^3_{03}&=&\frac12g^{33}\partial_t g_{33}&=&\phantom{-}\frac12(t^{-1}-W_t)\\[2mm]  
\Gamma^1_{33}&=&-\frac12g^{11}\partial_x g_{33}&=&\phantom{-}\frac12g^{11}  
W_x g_{33}\\[2mm] 
\Gamma^3_{13}&=&\frac12g^{33}\partial_x g_{33}&=&-\frac12W_x  
\end{array} 
\end{eqnarray*}

\section*{Acknowledgement} 
 
I am indebted to A.D.\ Rendall for suggesting me this problem and 
for useful discussions. 
 
\section*{Literature}  
  
%{\bf Berger, B.K.} (1974) Quantum graviton creation in a model  
%universe,  
%{\em Ann. Phys.}, {\em 83}: 458-490.  
  
{\bf Chru\'sciel, P.T., J.~Isenberg and V.~Moncrief} (1990)   
Strong cosmic censorship in polarised Gowdy spacetimes,  
{\em Class. Quant. Grav.}, {\em 7}: 1671-1680.  
  
{\bf Gowdy R.H.} (1974)   
Vacuum spacetimes with two-parameter spacelike isometry groups 
and compact invariant hypersurfaces: topology and boundary 
conditions,  
{\em Ann. Phys.}, {\em 83}: 203-241.  
 
%{\bf Hawking, S.W.} (1968) The existence of cosmic time 
%functions,  
%{\em Proc. Roy. Soc. Lond.}, {\em A 308}: 433-435.  
 
%{\bf Hawking, S.W. and R. Penrose} (1970) The singularities of  
%gravitaional collapse and cosmology,  
%{\em Proc. Roy. Soc. Lond.}, {\em A 314}: 529-548.  
  
%{\bf Hopf, H. and W. Rinow} (1931) "Uber den Begriff der  
%vollst"andigen differentialgeometrischen Fl"ache,  
%{\em Comm. Math. Helv.}, {\em 3}: 209-225.  
  
%{\bf Isenberg, J. and V.~Moncrief} (1982)   
%The existence of constant mean curvature foliations of Gowdy 
%3-torus spacetimes,  
%{\em Comm. Math. Phys.}, {\em 86}: 485-493.  
 
{\bf Isenberg, J. and V.~Moncrief} (1990)   
Asymptotic behavior of the gravitational field and the nature  
of singularities in Gowdy spacetimes,  
{\em Ann. Phys.}, {\em 199}: 84-122.  
   
{\bf Moncrief, V.} (1981)   
Global properties of Gowdy spacetimes with $\mathbb T^3\times\mathbb R$  
topology, {\em Ann. Phys.}, {\em 132}: 87-107.  
  
{\bf Rendall, A.D.} (2000)   
Fuchsian analysis of singularities in Gowdy spacetimes beyond analyticity,  
{\em Class. Quant. Grav.}, {\em 17}: 3305-3316.  
   
{\bf Ringstr\"om, H.} (2002)   
On Gowdy vacuum spacetimes,  
{\em preprint}, {\tt gr-qc/0204044}.  

%{\bf Seifert, H.-J.} (1977)   
%Smoothing and extending cosmic time functions,  
%{\em Gen. Rel. Grav.}, {\em 8}: 815-831.  
  
\end{document}